\documentclass[12pt]{iopart}
\usepackage{graphicx}
\usepackage{amssymb}
\usepackage{fontenc}
\usepackage{color}
\definecolor{Correction}{rgb}{0.0,0.0,0.0}  

\begin{document}

\newcommand{\ud}{\mathrm{d}}
\eqnobysec
\title{ PSI Effects on Plasma Burn-through in JET}
\author{Hyun-Tae Kim$^{1,2}$, A.C.C. Sips$^{1,3}$, W. Fundamenski$^{1}$, and EFDA-JET contributors*}
\address{JET-EFDA Culham Science Centre, Abingdon, OX14 3DB, UK.}
\address{$^1$Department of Physics, Imperial College London, Prince Consort Road, London, SW7~2AZ, UK}
\address{$^2$EURATOM/CCFE Fusion Association, Abingdon, OX14~3DB, UK}
\address{$^3$JET/EFDA, Culham Science Centre, Abingdon, OX14~3DB, UK}
\address{*See annex to F. Romanelli et al, Fusion Energy 2010  (Proc. 23rd IAEA Conf., Deajeon, 2010) IAEA Vienna.}
\ead{hyun.kim09@imperial.ac.uk}

\newpage

\begin{abstract}
Plasma Surface Interaction(PSI) effects on plasma burn-through are compared for the carbon wall and the ITER-Like Wall(ILW) at JET. For the carbon wall, the radiation barrier and $C^{2+}$ influx have a significant linear correlation whereas the radiation barrier in the ILW does not have such a linear correlation with $Be^{1+}$ influx. The JET data are explained by the simulation results of the DYON code. The radiation barrier in the carbon wall JET is dominated by the carbon radiation, but the radiation barrier in the ILW is mainly from the deuterium radiation rather than the beryllium radiation. 
\end{abstract}

\section{Introduction}

Tokamak start-up consists of the plasma break-down phase, the plasma burn-through phase, and the ramp-up phase of the plasma current $I_p$ \cite{ITERphysics1999}. 

The Townsend avalanche theory\cite{tanga}\cite{lloyd_D} is generally used to calculate the required electric field for plasma break-down at a given prefill gas pressure and effective connection length.  The minimum electric field for plasma break-down in ITER has also been calculated by using the Townsend criterion \cite{ITERphysics1999}. 

The Townsend criterion suggests the condition only for electron avalanche, i.e. the plasma break-down. In order for $I_p$ to increase in the $I_p$ ramp-up phase, sufficient ionization of prefill gas, the plasma burn-through, is required. The remaining neutrals result in significant electron power losses due to the radiation and ionizations, preventing the electron temperature from increasing in the $I_p$ ramp-up phase, which is necessary for $I_p$ to increase\cite{tanga}. Hence, the Townsend criterion is not sufficient to explain non-sustained breakdown discharges where $I_p$ does not increase after the plasma break-down.

The minimum loop voltage for plasma burn-through of the prefill gas, the burn-through criterion, is generally higher than that calculated using the Townsend criterion\cite{ITERphysics1999}. Therefore, tokamak operation space must be determined considering the requirements for full ionization, burn-through criterion. In the 2009 JET campaign, more than 100 shots  failed during the burn-through phase. These burn-through failures can be prevented by understanding the key aspect of the plasma burn-through physics. Furthermore, the allowable toroidal electric field for ITER start-up is limited up to $0.3[V/m]$ due to the engineering issues resulting from the use of superconducting poloidal coils and a continuous vacuum vessel\cite{ITERphysics2007}. For reliable start-up using a low electric field, ECH-assisted start-up is planned in ITER\cite{ITERphysics2007}. A more accurate estimation of the ECH power can be obtained by better understanding the burn-through criterion.
 
In order to achieve the required degree of ionization during the plasma burn-through phase, the ohmic heating power must exceed the maximum of the total electron power loss. The electron power loss in the burn-through phase is mainly due to the radiation and ionization power losses, so that the peak of the total electron power loss is also dominated by the radiation and ionization power losses. \textcolor{Correction}{The ionization power loss changes in a way analogous to the radiation power loss since both power losses are functions of the product of electron and deuterium atom densities, $n_e n^0_d$. The radiation barrier, which is defined as the maximum of the radiation power loss during the plasma burn-through phase, is directly measurable using bolometry. Hence, the radiation barrier is very useful to estimate the peak of total electon power loss, thereby determining the burn-through criterion, i.e. the minimum loop voltage required for plasma burn-through.} It is generally known that not only the burn-through of the prefill gas(Deuterium) but also the burn-through of the impurities from the first wall is important since the impurities can result in significant radiation power loss until they are fully ionized\cite{ITERphysics1999}. Hence, in this article, the effects of impurity influx on the radiation barrier in the carbon wall JET and the beryllium wall JET are compared. 

The effect of impurities on plasma burn-through in ITER has been simulated assuming a constant content of carbon and beryllium\cite{lloyd_ITER}. However, the treatment of impurity in the simulation was overly simplified. In order to simulate the impurity effects in the burn-through phase, the evolution of impurity densities should be calculated considering Plasma Surface Interaction(PSI) effects. The new burn-through model including the PSI effects, used in the DYON code, has been validated in JET\cite{DYON}. The JET data in the carbon wall JET and ITER-Like Wall(ILW) JET are compared with the simulation results of the DYON code. 

The structure of this paper is following. In order to calculate the impurity influx, we need to know the electron temperature $T_e$ since the inverse photon efficiency is a function of $T_e$. The $T_e$ at the peak of a specific line emission can be obtained by using the fractional abundance in non-coronal equilibrium. In section 2.1, the details about this method is explained. In section 2.2, the correlation between the impurity influx calculated by using the obtained $T_e$ and the radiation barrier in JET is presented. In section 2.3, the JET data is explained by the simulation results of the DYON code. In section 3, conclusions are presented. 

\section{PSI Effects on Plasma Burn-through}  
\subsection{Fractional Abundance of Impurity in Non-coronal Equilibrium} 
The impurity influx $\Gamma_I^{z+}\textcolor{Correction}{[m^{-2} s^{-1}]}$ can be calculated using a specific line emission from the impurity, $I^{\lambda[nm]}\textcolor{Correction}{[photons \ m^{-2} s^{-1} ]}$, and the corresponding inverse photon efficiency $SXB(T_e)$, 
\begin{eqnarray}
SXB(T_e) = \frac{<\sigma v_{iz}>}{b_r <\sigma v_{exc, lm}>}
\end{eqnarray}
where $<\sigma v_{iz}>$ and $<\sigma v_{exc, lm}>$ are the ionization rate coefficient and the excitation rate coefficient for transition from state $l$ to $m$, resulting in the subsequent release of a specific line emission, and $b_r$ is the branching ratio for the particular optical transition, i.e. SXB = ionizations/photon. Hence, the particle influx into the charge state can be calculated by the photomutiplier tube data measuring a specific line emission, i.e. $\Gamma_I^{z+}\textcolor{Correction}{[m^{-2} s^{-1}]}= I^{\lambda[nm]}\textcolor{Correction}{[photons \ m^{-2} s^{-1} ]} \times SXB^{\lambda[nm]} (T_e)$\cite{stangeby}. \textcolor{Correction}{The photomultiplier tube data means the number of photons, line integrated along a line of sight. In this article, the averaged value of the photomultiplier tube data measured by two orthogonal lines of sight, i.e. vertical and horizontal lines of sight, is used for the impurity influx calculation.} The values of SXB used in this article are adopted from the Atomic Data and Analysis Structure(ADAS) package\cite{Summers}. 

In order to calculate the impurity influx at a specific moment, the corresponding electron temperature $T_e$ is required as SXB is a function of $T_e$. However, the measurement of $T_e$ during the burn-through phase is not accurate due to the significant diagnostic errors in this phase. During the plasma burn-through phase, the dominant charge state of the impurity rises as $T_e$ increases. This results in a maximum in time(peak) of a specific line emission of the impurity. In coronal equilibrium, the fractional abundance of the charge state of the impurity is determined by $T_e$. Hence, the corresponding $T_e$ at the peak of the photon emission can be obtained using the fractional abundance. However, in the case of the plasma burn-through phase, coronal equilibrium is not valid due to the significant particle transport along the open magnetic field lines. In order to estimate the correct $T_e$ at the peak of the line emission, the fractional abundance should be calculated by the particle balance of each charge state including particle transport.  \textcolor{Correction}{For this calculation, we assumes that all neutrals are backscattered and ions are recyclied to neutrals at the wall with 1 recycling coefficient. According to this assumption, neutral influx is only from ion recycling, and the particle balances can be simplified as shown below.} 
\begin{eqnarray}
\fl 0=-R_{I,iz}^{0} n_I^{0}+R_{I,rec}^{1+ } n_I^{1+}+\sum_z \frac{n_I^{z+}}{n_e\tau_p} \nonumber \\
\fl 0=R_{I,iz}^{(z-1)+ } n_I^{(z-1)+}-R_{I,iz}^{z+ } n_I^{z+}+R_{I,rec}^{(z+1)+ } n_I^{(z+1)+}-R_{I,rec}^{z+ } n_I^{z+}-\frac{n_I^{z+}}{n_e\tau_p} \label{fractionalabundance}
\end{eqnarray}
where $\tau_p$ is the particle confinement time for ions, and $R_{I,iz}^{z+ }$ and $R_{I,rec}^{z+ }$ indicate the rate coefficients for ionization and recombination, respectively. The confinement time $\tau_p$ during the burn-through phase can be approximately calculated as shown below.
\begin{eqnarray}
\tau_p[sec]= \frac{L_f[m]}{C_s[m/sec]} 
\end{eqnarray}
where $L_f$ is an effective connection length\cite{tanga},
\begin{eqnarray}
L_f[m]=0.25 \times a[m] \times \frac{B_{\phi}[T]}{B_{\perp}[T]},
\end{eqnarray} 
and $C_s$ is the sound speed,
\begin{eqnarray}
C_s[m/sec] = \sqrt{\frac{eT_e[eV]}{m_D[kg]}}.
\end{eqnarray}
$m_D$ is the mass of deuterium and $e$ is a unit charge. In the case of the burn-through phase in JET, we can assume that the minor radius $a=0.8[m]$, the toroidal magnetic field $B_{\phi}=2.3[T]$, the stray magnetic field $B_{\perp}=10^{-3}[T]$, and $T_e = 5 \sim 10[eV]$. The resultant $\tau_p$ is $29[msec]$ when $T_e=5[eV]$ and $21[msec]$ when $T_e = 10[eV]$. According to this, it can be justified that $\tau_p$ during the burn-through phase in JET is between $10$ and $50[msec]$.

\begin{figure}
\begin{center}
\includegraphics[width=1\textwidth]{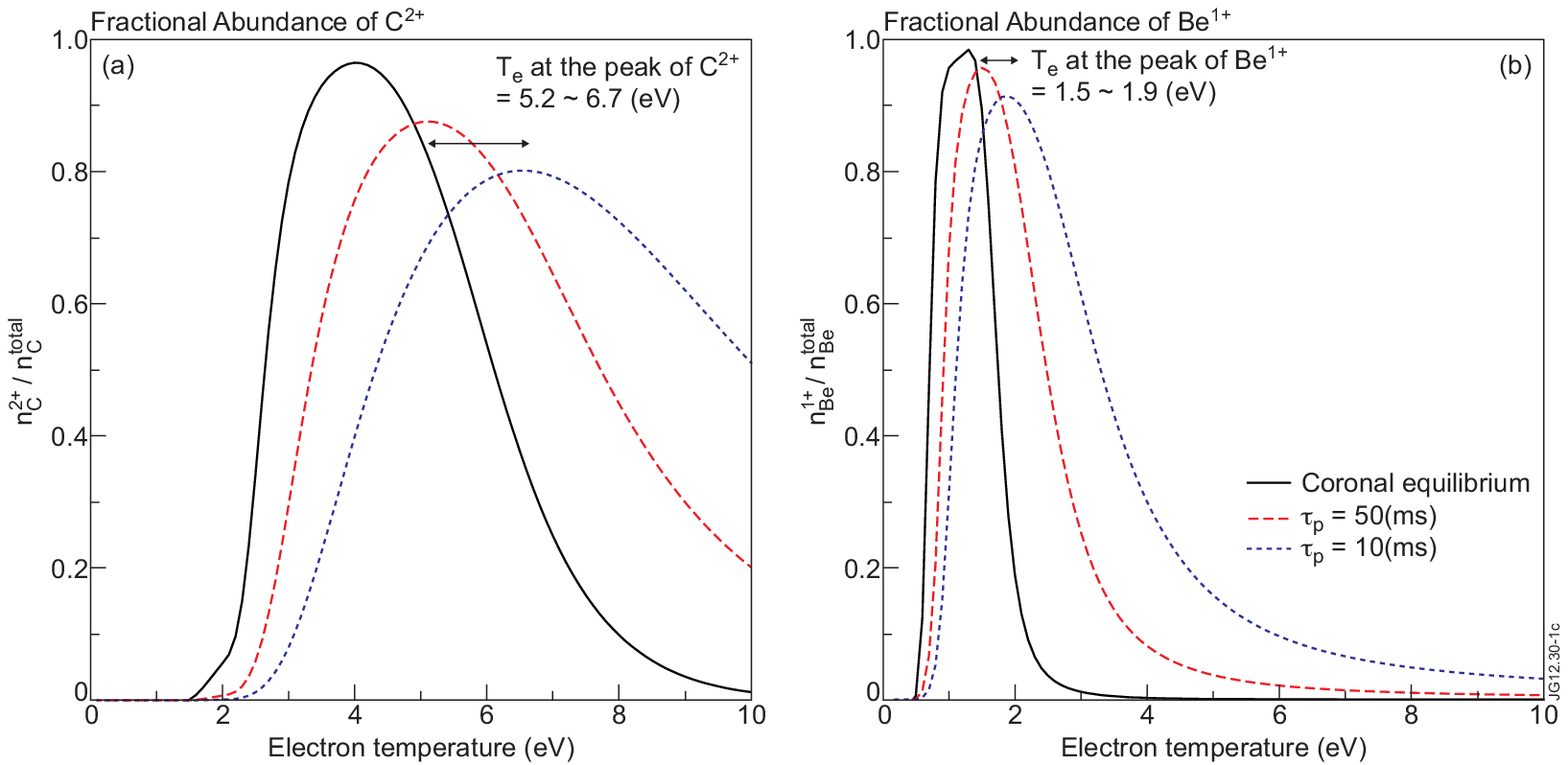} 
\caption{(a) and (b) show the fractional abundance of $C^{2+}$ and $Be^{1+}$, respectively. In both figures, each line indicates the assumed equilibrium or particle confinement time : solid black(Coronal equilibrium), dashed red($\tau_p=50[ms]$), and dashed blue($\tau_p=10[ms]$).} \label{PSI_Figure1}
\end{center}
\end{figure}

Figure \ref{PSI_Figure1} shows the fractional abundances of $C^{2+}$ and $Be^{1+}$ for $\tau_p$ values of $10[msec]$ and $50[msecs]$. Compared to the case of coronal equilibrium($\tau_p = \infty$), the peaks of $C^{2+}$ and $Be^{1+}$ are shifted to higher $T_e$ due to the transport effect. According to Figure \ref{PSI_Figure1}, the range of $T_e$ at the peak of $C^{2+}$ and $Be^{1+}$ are $5.2 \sim 6.7[eV]$ and $1.5 \sim 1.9[eV]$, respectively. 

\subsection{Radiation Barrier versus Impurity Influx in JET } 

\begin{figure}
\begin{center}
\includegraphics[width=1\textwidth]{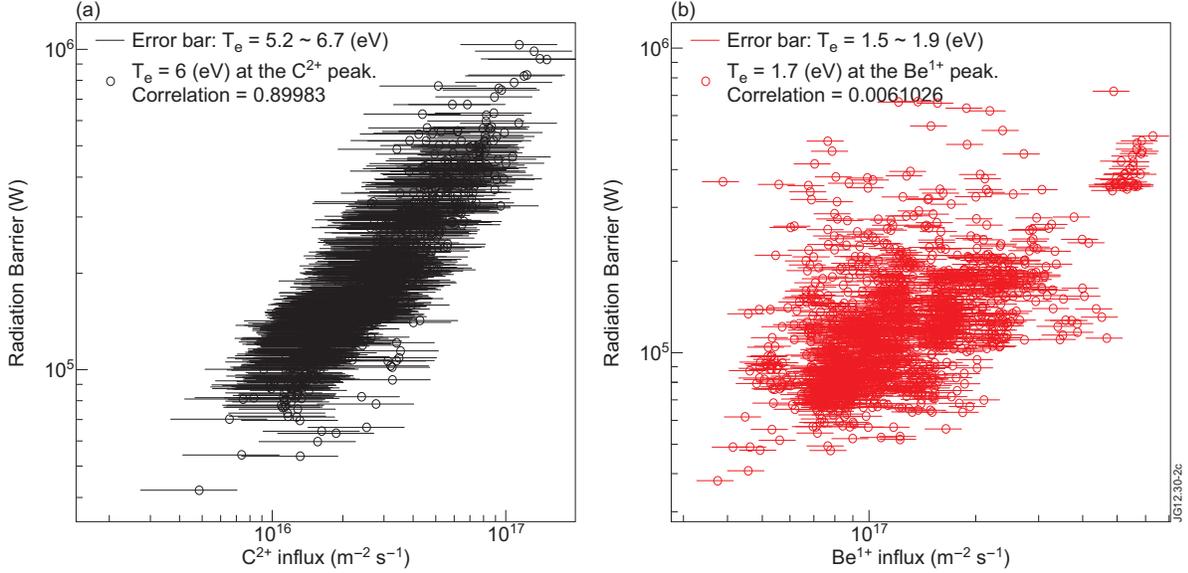}
\caption{(a) shows the radiation barrier at different $C^{2+}$ influx calculated by the $T_e(t_{C^{2+} peak}) = 6[eV]$ in the carbon wall JET. The error bars of $C^{2+}$ influx are for $T_e(t_{C^{2+} peak}) = 5.2$ or  $6.7[eV]$. (b) indicates the radiation barrier at different $Be^{1+}$ influx calculated by  the $T_e(t_{Be^{1+} peak}) = 1.7[eV]$ in the ILW. The error bars of $Be^{1+}$ influx are for $T_e(t_{Be^{1+} peak}) = 1.5$ or  $1.9[eV]$. The linear correlation coefficients for the carbon wall and the ILW are $0.89983$ and $0.0061026$, respectively.} \label{PSI_Figure2}
\end{center}
\end{figure}

The influx of $C^{2+}$ and $Be^{1+}$ are calculated by using photomultiplier tube data ($465[nm]$ of $C^{2+}$  and $527[nm]$ of $Be^{1+}$) and SXB values assuming $T_e$ at the each peak of the line emission is $6[eV]$ and $1.7[eV]$, respectively, as shown below. 
\begin{eqnarray}
\Gamma_C^{2+}[m^{-2} sec^{-1}] = I^{465[nm]}\textcolor{Correction}{[photons \ m^{-2} s^{-1} ]} \times SXB^{465[nm]}(6[eV]) \\ \nonumber
\Gamma_{Be}^{1+}[m^{-2} sec^{-1}] =I^{527[nm]}\textcolor{Correction}{[photons \ m^{-2} s^{-1} ]} \times SXB^{527[nm]}(1.7[eV])
\end{eqnarray}
In this calculation, $n_e$ is  assumed to be $10^{18}[m^{-3}]$ since the dependence of SXB on $n_e$ is small enough to be ignored. The calculated impurity influx and the radiation barrier measured by bolometry in JET are presented in Figure \ref{PSI_Figure2}. The error bars in Figure \ref{PSI_Figure2} correspond to the range of $T_e$ obtained in Figure \ref{PSI_Figure1}.  

It should be noted that the linear correlation coefficient in the carbon wall JET is $0.9$  while it is only $0.0061$ for the ILW. This implies that the radiation barrier  was strongly affected by the impurity influx in the carbon wall JET, but the effect of impurity is not important in the ILW.  

\subsection{Simulation results of the DYON code} 

\begin{figure}
\begin{center}
\includegraphics[width=1\textwidth]{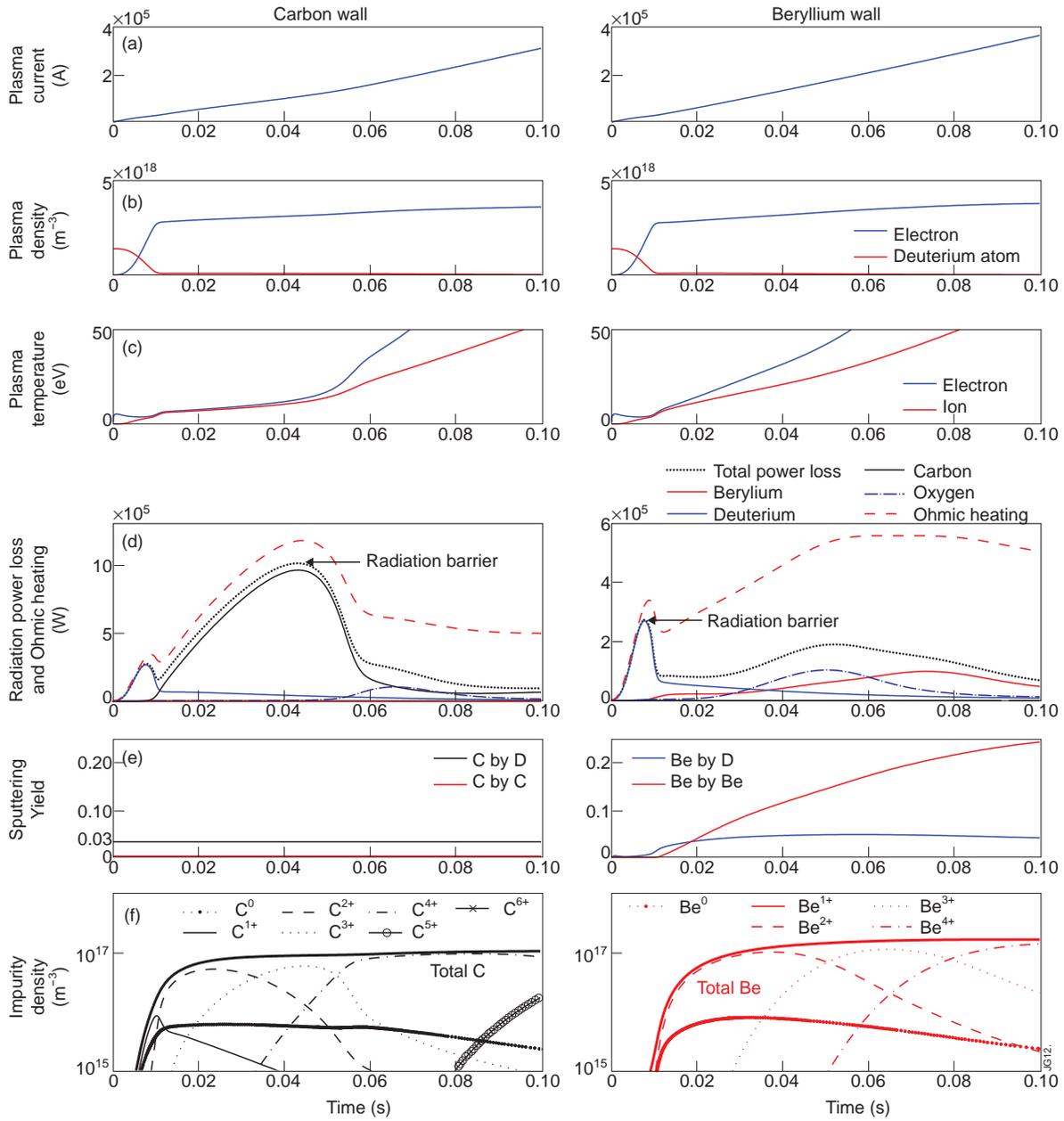}
\caption{Simulation results in the carbon wall and the ILW under the identical conditions are presented: (a)Plasma current, (b)electron density(blue) and deuterium atom density(red), (c)electron temperature(blue) and ion temperature(red), (d)radiation power losses \textcolor{Correction}{and ohmic heating power}(dashed black: total radiation power loss, solid blue: deuterium radiation, solid red: beryllium radiation, solid black: carbon radiation, dashed blue: oxygen radiation, \textcolor{Correction}{and dashed red: ohmic heating power}), (e)sputtering yield(solid black: carbon sputtering due to incident deuterium ion, solid blue: beryllium sputtering due to incident deuterium ion, solid red: self-sputtering yield), (f)impurity densities in each charge state.} \label{PSI_Figure3}
\end{center}
\end{figure}

In order to investigate the PSI effects on the radiation barriers, plasma burn-through in the carbon wall  and ILW is simulated using the DYON code\cite{DYON}. The identical conditions(prefill gas pressure $= 5\times 10^{-5}[Torr]$, loop voltage $= 25[V]$) are given for the simulations except the wall sputtering models. For carbon wall, chemical sputtering is dominant when incident deuterium ion energy is lower than 100[eV]\cite{Mech1998}. Since chemical sputtering yield is not subject to an incident ion energy, the sputtering yield in carbon wall is assumed to be constant, i.e. $Y^D_C=0.03$ and $Y^C_C = 0$. The details of the PSI models are given in \cite{DYON}. In the case of beryllium wall, the PSI effect is dominated by physical sputtering\cite{Stamp}. The formula for physical sputtering is given in \cite{sputteringformula}\cite{surfacebindingenergy}.  The physical sputtering for the simulation is modeled as a function of $T_e$ and $T_i$. Figure \ref{PSI_Figure3} shows the simulation results in the carbon wall JET(left) and the ILW(right). The sputtering yields in the simulation are shown in Figure \ref{PSI_Figure3}(e). 

\textcolor{Correction}{As shown in Figure \ref{PSI_Figure3}(d), ohmic heating power is comparable to the total radiation power loss until the radiation barrier is overcome.} Consistently, it is not until the radiation barrier is overcome that $T_e$ begins to increase, since the significant radiated power losses impede $T_e$ from increasing. As can be seen in Figure \ref{PSI_Figure3}(c), it is after $0.05[sec]$ and $0.01[sec]$ that $T_e$ rises steeply in the carbon wall and the beryllium wall, respectively. This implies that the corresponding radiation barrier are located at $0.05[sec]$ and $0.01[sec]$ as indicated in Figure \ref{PSI_Figure3}(d).  The radiation barrier in the carbon wall JET is dominated by the carbon radiation(solid black) whereas it is mainly from the deuterium radiation(solid blue) in the beryllium wall rather than the beryllium radiation(solid red). In other words, the radiation barrier in the ILW is not dependent on the beryllium content whereas it does depend on the carbon for the carbon wall. It should be noted that this simulation result is consistent with the results presented in Figure \ref{PSI_Figure2}. 

The maximum radiation from carbon is about 10 times higher than that of beryllium in Figure \ref{PSI_Figure3}(d), while $n_e$ and total impurity content do not differ much as shown in Figure \ref{PSI_Figure3}(b)and \ref{PSI_Figure3}(f). According to this, it can be seen that the significant discrepancy in the radiation power losses results from the different radiation power coefficients of carbon and beryllium during the burn-through phase.     

\section{Conclusion}

The influx of $C^{2+}$ and $Be^{1+}$ during the plasma burn-through phase are calculated assuming non-coronal equilibrium at the each peak of the line emission of the impurities. The calculated impurity influx in the carbon wall has a strong linear correlation with the radiation barrier, but such a correlation does not appear in the ILW at JET. This result is explained with the simulation results of the DYON code. The radiation barrier in carbon wall is dominated by the carbon radiation. However, in the ILW, the deuterium radiation is dominant in the radiation barrier. Hence, for the ILW, the PSI effects do not seriously influence on the plasma burn-through as in the carbon wall. This implies that the required ohmic heating power for plasma burn-through will be lower in the ILW compared to the carbon wall in cases where the prefill gas pressures are identical. 

\section*{Acknowledgment}
This research was funded partly by the Kwanjeong Educational Foundation and by the European Communities under the contract of Association between EURATOM and CCFE. The views and opinions expressed herein do not necessarily reflect those of the European Commission. This work was carried out within the framework of the European Fusion Development Agreement.

\section*{References}

\bibliography{referenceKHT}
\bibliographystyle{unsrt}

\newpage

\end{document}